\documentclass[conference,onecolumn,12pt]{IEEEtran}
\usepackage{cite}
\usepackage{graphicx}
\usepackage{epstopdf}
\ifCLASSOPTIONcompsoc
  \usepackage[caption=false,font=normalsize,labelfont=sf,textfont=sf]{subfig}
\else
  \usepackage[caption=false,font=footnotesize]{subfig}
\fi
\usepackage{amsmath}
\usepackage{amssymb}
\usepackage{multirow}
\usepackage[ruled,lined,boxed]{algorithm2e}

\usepackage{array}
\newcolumntype{M}[1]{>{\raggedright}m{#1}}
\usepackage{stfloats}
\usepackage[hyphens]{url}

\makeatletter
\def\ps@IEEEtitlepagestyle{%
  \def\@oddfoot{\mycopyrightnotice}%
  \def\@evenfoot{}%
}
\def\mycopyrightnotice{%
  {\footnotesize Accepted in IEEE Symposium on Computers and Communications June, 2016\hfill}
  \gdef\mycopyrightnotice{}
}
\begin{document}

\title{Spectrum Resource Management and Interference Mitigation for D2D Communications with Awareness of BER Constraint in mmWave 5G Underlay Network}

\author{\IEEEauthorblockN{Zeineb Guizani and
Noureddine Hamdi}
\IEEEauthorblockA{
Communications Systems Laboratory, National Engineering School of Tunis,\\
University of Tunis El Manar, Tunisia \\
}
}

\maketitle

\begin{abstract}
The work presented in this paper deals with the issue of massive demands for higher capacity. For that matter, we investigate the spectrum resource management in outdoor mmWave cell for the uplink of cellular and D2D communications. Indeed, we provide a first insight how to optimize the system performance in terms of achievable throughput while realizing a compromise between the large number of admitted devices and the generated interference constraint. We propose a mathematical formulation of the optimization objective which falls in the mixed integer-real optimization scheme. To overcome its complexity, we apply a heuristic algorithm and test its efficiency through simulation results with a particular regard to the BER impact in the QoS. 
\end{abstract}

\IEEEpeerreviewmaketitle
\section{Introduction}
The continual evolution of the technological era has advanced drastically making it so far difficult to gauge precisely where it is headed in the future. Thus, key global fifth Generation (5G) players from around the world are working together to clear the vision and the strategic orientation of the standards. Still, it is undoubted that nowadays communication converges to a new design where device-centric communication horn in human-centric communication as a direct result of the connected devices explosion.

The network function virtualization is an enabling technology that performs abstraction of physical resources by the means of slicing. Combined with the software defined networking, it fosters the development and the diversification of services provided by telecommunication operators. By introducing the massive MIMO, hundreds and thousands of antennas achieve coherent and highly precise transmission and reception leading to substantial gains in capacity and energy efficiency \cite{survey5G2016}.

Moreover, revolutionary inventions have made possible the exploitation of the millimeter-wave bands ranging between 11 and 300 GHz. It is considered as an effective solution in the struggle against the spectrum shortage. This is because it managed to simplify the network design due to the wavelength shortness, improve the quality of wireless transmission as well as interference mitigation. For that, extensive measures for channel modeling have been carried out for different frequencies to pave the way for new algorithms and protocols to provide multi-gigabit services. To maximize the system throughput and ensure fairness among users, dynamic scheduling and congestion control scheme was proposed in \cite{mmWmultihop} that exploits the benefits of the millimeter wave (mmWave) in 28 GHz band and boosts its performance by using the multihop relaying technology.

One of the main design concerns of realizing the separation of control and data plane in network architecture is the Heterogeneous Cloud Radio Access Network (H-CRAN). As indicated by its name, the baseband processing and the network control are shifted to a centralized zone denoted base band unit (BBU) pool. Environmental friendly, it has the capability to deal with very complex heterogenous structures and enablers in a low cost and effective way. Another particularity of 5G system is their intense heterogeneity (heterogeneous networks) in terms of transmit powers, supported frequency bands, and of course the coexistence of cellular and peer-to-peer communications (i.e mobile-to-mobile (M2M) and device-to-device D2D communication for both underlay and overlay mode non orthogonal/orthogonal spectrum sharing with cellular users)\cite{M2MNajjar}.

Particularly, interdisciplinary research efforts have been carried to exploit the D2D communication advantage in spectrum reuse, network offloading, the massive access and the enhancement in user and cell throughput under the aggregated interference issue. This subject has been tackled in different directions. The majority of research studies prioritize cellular users over D2D users \cite{D2DqosSurvey}. The strategy adopted in \cite{RateQos} for example guaranties a required rate for cellular users preserving hence a satisfactory quality of service (QoS). This was done through global mechanisms of power control and resource allocation. Even if the D2D communication needs less requirements compared to conventional users, it is necessary to protect them from the superposed interference and make them meet their target signal to interference plus noise ratio (SINR) as well as improve their data rates. Therefore, interference mitigation schemes and different strategies of mode selection are used to ensure the symbiotic coexistence as proposed in \cite{QoSD2D, mixedMode}.

There is a great deal of concern in choosing between the centralized and decentralized approaches as surveyed in \cite{D2DsurveyEkram}.
In the works where the centralized scheme is adopted, the base station (BS) fully manages the radio resources according to the channel state information and the traffic demand then decides of the scheduling. Even if this choice alleviates the interference, it presents a main drawback that is the massive network signaling for control. In the decentralized strategy, instead, the D2D equipments communicate and share resources with cellular users autonomously which decreases the overhead but generates less interference management \cite{energyWang, CentraliDecentakizD2D}.

There have been many challenging works to ensure the critical trade-off between the massive connectivity and the interference issues. This is enabled through two approaches. The first allows the D2D users to reuse the spectrum resources from more than one cellular user terminal as investigated in \cite{D2DmultiReuse,D2DmultiReuse2}. The second by allowing cellular users to share their resources by several D2D users as analyzed in \cite{energyWang,multisharing}. However, this approach is challenging and scarcely investigated especially in real scenarios where the number of D2D users is high. So, the majority of works simplify it by allowing cellular users to share their spectral resources with at most two D2D users. Such strategy is used to reduce the interference at the BS and to fulfill the QoS of cellular users.

An increasing interest in using D2D communications for underlay cellular network in mmWave bands has been experienced. The work presented in \cite{mmWvCroosLayer} combines both H-CRAN and SDN technologies to provide an efficient scheduling for indoor environment. Besides, it resorts to the cross layer design for a global control of physical (PHY), medium access control (MAC) and Network layers. The work of \cite{mmWavSmallCell} proposed the D2DMAC; a scheduling algorithm for both access and backhaul; in 60 Ghz mmWave band for small cells. In this scheme, where a centralized control is adopted, the spatial reuse gain is achieved through the concurrent transmission and the priority of D2D users is ensured through the path selection. For widespread adoption of D2D communication in underlay cellular network for the millimeter wave spectrum band range is undoubtable. Nevertheless, the most foreseen applications are for the indoor and relatively rare are the works that investigate it in the outdoor.

Motivated by the above facts, the present paper deals with the spectrum reuse issue in mmWave 28GHz band for an outdoor scenario wherein the number of D2D pairs largely exceeds cellular users. We develop a radio resource management scheme that aims to improve the network performance in terms of achievable data rate for both cellular and D2D users through spatial and multi-user gain combined with efficient interference policy. A particular regard is dedicated to the role that plays the BER in the QoS transmission.

The paper is organized as follows. Section II introduces the system description and assumption. Section III deals with the mathematical formulation of the optimization problem. In section IV, near-optimal solution algorithm for resource reuse is provided. Section V unleashes the simulations results. Finally, section V summarizes the achieved work.

\section{System Description and Assumption}
\subsection{Network Model}
\begin{figure}
\centering
\includegraphics[width=8cm]{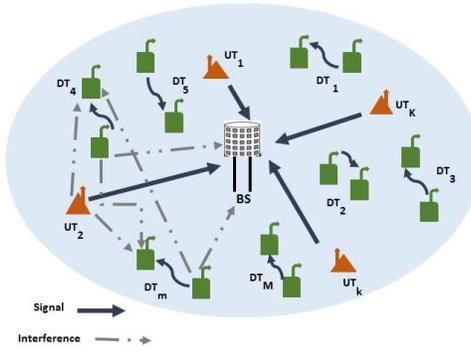}
\caption{System model with illustration of the interference between $UT_2$ and $DT_4$ and $DT_M$ with whom he shares its RB. }\label{SystemModel3}
\end{figure}
The D2D communication enabled single cell environment, adopted for study in this paper, is presented in Fig.\ref{SystemModel3}. The signal coverage is ensured by a BS localized at the center that serves in the uplink mode $K$ cellular user terminals (UT) labelled $\mathbf{S}=\{s_{1},s_{2},...,s_{K}\}$. These UTs share the same radio resources with $M$ device terminal (DT) pairs labelled $\mathbf{M}=\{d_1, d_2,...,d_M\}$ on the underlay mode. The DT transmitter is denoted by $d_{TX}$ and the receiver by $d_{RX}$. A binary indicator $\rho_{d,s}$ is set to 1 when $UT_s$ shares its resource block (RB) with a $DT_d$. Note that DT pairs are allowed to reuse the RB of only one UT, contrarily to the UTs who can share their RBs to many DTs as long as their channel status allows them.
The received signal at the BS can be written as:
\begin{equation} \label{eq1}
\begin{aligned}
{y_B}&=\sqrt{P_s}h_{s,B}x_s+n_{s,B}
&+\underset{d=1} {\overset{M}{\sum}} \rho_{d,s}(\sqrt{P_d}h_{d_{TX},B}x_d+n_{d,B})
\end{aligned}
\end{equation}
where $P_s$ and $P_d$, $x_s$ and $x_d$ are the $UT_s$ and $DT_d$ transmit power and transmitted data. $h_{X,Y}$ and $n_{X,Y}$ denote the channel $X-Y$ transfer function and the additive white gaussian noise (AWGN) power. \\
Likewise, we define the received signal at the $d_{RX}$ by:
\begin{equation} \label{eq2}
\begin{aligned}
y_{d_{RX}}
&=\sqrt{P_d}h_{d_{TX},d_{RX}}x_d+n_{d,d}\\&+ \underset{s=1} {\overset{K}{\sum}} \rho_{d,s}(\sqrt{P_s}h_{s,d_{RX}}x_s+n_{s,d})\\
&+\underset{d^{'}\in \mathbb{M}\setminus \{d\}} {\overset{M}{\sum}}\underset{s=1} {\overset{K}{\sum}} \rho_{d^{'},s}(\sqrt{P_{{d^{'}}_{TX}}}h_{{d^{'}_{TX}},d_{RX}}x_{d^{'}}+n_{{d^{'}},d})
\end{aligned}
\end{equation}
where $d^{'}$ denotes another DT which reuses the RB of the same $UT_s$ as $DT_d$.
\subsection{Radio propagation model}
The most studied bands in the millimeter wave are the 28, 38, 60, 71-76 and 81-86 GHz. Results show that they provide ubiquitous throughput, high quality of wireless links, massive antenna deployment and clear network design. However, these mmWave bands are extremely directive and usually subject to signal attenuation due to obstacles and atmospheric absorbtion. Tab.\ref{mmWattenuation} gives an overview how these frequency bands are affected by rain attenuation and oxygen absorbtion. The use of techniques such as beamforming and directional antennas has helped to address these challenges.
\begin{table}[!t]
\caption{The propagation characteristics of mmWave communications in different bands\cite{mmWlimits}} \label{mmWattenuation}
\center
\begin{center}
\begin{tabular}{|c|c|c|c|}
  \hline
    \multirow{2}{1.5cm}{Frequency  band (GHz)}& \multicolumn{2}{|c|}{Rain attenuation for 200 m }&\multirow{2}{2.3cm}{Oxygen absorbtion at 200 m (dB)} \\ \cline{2-3}
  &5 mm/h (dB)& 25 mm/h (dB)& \\ \hline
  28 & 0.18 & 0.9 & 0.04 \\\hline
  38 & 0.26 & 1.4 & 0.03 \\\hline
  60 & 0.44 & 2 & 3.2 \\\hline
  73 & 0.6 & 2.4 & 0.09 \\\hline
  \hline
\end{tabular}
\end{center}
\end{table}

In this work, we consider 28 GHz mmWave band. Realistic outdoor propagation conditions for 28 GHz mmWave are proposed by \cite{73PLmodel}. They were collected through large scale measurements which are carried out in New York city. Famous for its very dense environment of users and obstacles, the line of sight is practically improbable in such locations. It is characterized by angular signal copies with different delays \cite{mmWvErkip}.
Further, the path loss (PL) model separates the Line of sight (LOS) and the non line of sight (NLOS) components and associates to each of them the corresponding shadowing. The PL is calculated as $PL=PL_{LOS}+PL_{NLOS}$. For a distance $d$, each component is considered as $PL_{X}(d)[dB]=\mu+10\nu \log10[d(m)]+\xi, \ \xi \sim \mathbf{N}(0,\sigma^2)$ with $\mu$ as the PL coefficient, $\nu$ as its exponent and $\xi$ as its corresponding lognormal shadowing with mean 0 and variance $\sigma^2$. 
We insert probability to the lognormal path-loss and shadowing model. By that, we favorite DTs to receive more LOS signals given their close TX-RX proximity. The D2D link PL is calculated as in (\ref{eq3}):
\begin{equation} \label{eq3}
\begin{aligned}
PL1=p1\ PL_{LOS}+(1-p1)\ PL_{NLOS}
\end{aligned}
\end{equation}
For the rest of links, i.e, (BS-D2D), (D2D-UT) or (UT-UT) and (UT-BS) the PL is expressed as follows:
\begin{equation} \label{eq4}
\begin{aligned}
PL2=p2\ PL_{LOS}+(1-p2)\ PL_{NLOS}
\end{aligned}
\end{equation}
Moreover, works in \cite{Rice1} indicate that the multipath fading is likely to be a Rician channel rather than Rayleigh.
\subsection{Achievable data rate}
In this single cell network, it is assumed that the number of DTs largely exceeds the number of cellular users (UTs). Moreover, the bandwidth is divided into $N$ RBs of bandwidth $B_{RB}$ where $(N \geq K)$. Given the fact that we consider a fully loaded scenario, $V$ DTs are selected to be treated as cellular users who share their RBs with the remaining DTs in the case where $N>K$. Therefore, the set $\mathbf{S}$ is extended to include the $V$ elements. Besides, it is indexed $UT_s$ and referred to $DT_d$ such as $s=d^{*}$. It is worthy of noticing that a perfect channel state information is ensured at the base station and the inter-cell interference is well mitigated.

Owner of a RB in the set $\mathbf{S}$, each transmitter is allowed to share its spectrum resource with DTs if it satisfies the throughput-BER compromise that depends on the SINR. It is denoted $\gamma_s$ and can be written as:
\begin{equation} \label{eq5}
\begin{aligned}
\gamma_s=\frac{P_s (\alpha_s H_{s,B} +\beta_s H_{s,s}) }
{ \underset{d=1} {\overset{M}{\sum}} \rho_{d,s} P_d(\alpha_s H_{d_{TX},B} +\beta_s H_{d_{TX},s} )+N_s}\ \\
\geq {\gamma_s}^{th},\  s=1..N\ \ \ \ \ \ \ \ \ \ \ \ \ \ \ \ \ \ \ \ \ \ \ \ \ \ \ \ \ \
\end{aligned}
\end{equation}
In above, $\alpha_s$ and $\beta_s$ are opposite binary indicators which are used to differentiate between the UTs and DTs owners of RBs in $\mathbf{S}$. $(\alpha_s,\beta_s)$ equals (1,0) when $s\in [1,N-V]$ and $(\alpha_s,\beta_s)$ equals (0,1) otherwise. $H$ is the channel gain, $N_s$ is the noise power and ${\gamma_s}^{th}$ is the SINR threshold corresponding to $UT_s$.\\
The DTs are also under SINR requirements defined by the throughput-BER compromise. It is denoted $\gamma_d$ that must be greater than a threshold ${\gamma_d}^{th}$ and can be identified as:
\begin{equation} \label{eq6}
\begin{aligned}
\frac{P_d H_{d_{TX},d_{RX}}}{ \underset{s=1} {\overset{N}{\sum}} \rho_{d,s}P_s H_{s,d_{RX}} +\underset{s=1} {\overset{N}{\sum}} \underset{d^{'}\in \mathbb{M}\setminus \{d\}} {\overset{M}{\sum}} \rho_{d^{'},s} P_{{d^{'}}_{TX}} H_{{d^{'}}_{TX},d_{RX}} +N_d}\\= \gamma_d \geq {\gamma_d}^{th},\ \ \ \   d=1..M,\ \ \ \ \ \ \ \ \ \ \ \ \ \ \ \ \ \ \ \ \ \ \ \ \ \ \ \ \ \
\end{aligned}
\end{equation}
The data rates for $UT_s$ and $DT_d$ are given by
\begin{equation} \label{eq7}
\begin{aligned}
R_s= B_{RB}\log_{2}(1+ cst_s \gamma_s) \ \ \ \  \ \ \ \ \
\end{aligned}
\end{equation}
\begin{equation} \label{eq8}
\begin{aligned}
R_d= \underset{s=1} {\overset{N}{\sum}} \rho_{d,s} B_{RB}\log_{2}(1+ cst_d \gamma_d)
\end{aligned}
\end{equation}
where, $cst_d = -1.5/ \ln(5BER_d)$ and $cst_s = -1.5/ \ln(5BER_s)$. Each of them identifies the QoS imposed by a minimum probability of error, i.e, Bit Error Rate (BER).
\section{Mathematical Analysis}
In this mmWave outdoor network design, the sum-rate of UTs and DTs is considered as the objective function to maximize. The optimal solution for this resource sharing is to specify the admitted DTs per each UT (i.e a matrix $\rho=[N;M]$), where both BER-aware SINR requirements and a large number of admitted D2D users are respected.
The resource sharing optimization problem can be formulated as follows:
\begin{equation} \label{eq9}
\begin{aligned}
R_{Tot}= \underset{\rho}{\max}\{\underset{s=1}{\overset{N}{\sum}}  R_s + \underset{d=1}{\overset{M}{\sum}} R_d\}\
= \underset{\rho}{\max}\{ \underset{s=1}{\overset{N}{\sum}} B_{RB}\ \ \ \ \ \ \ \ \ \ \ \\
\log_{2}(1+cst_s  \times \frac{P_s (\alpha_s H_{s,B} +\beta_s H_{s,s}) }
{ \underset{d=1} {\overset{M}{\sum}} \rho_{d,s} P_d(\alpha_s H_{d_{TX},B} +\beta_s H_{d_{TX},s} )+N_s})\ \ \\
+ \underset{d=1}{\overset{M}{\sum}} \underset{s=1} {\overset{N}{\sum}} \rho_{d,s} B_{RB} \log_{2}(1+cst_d \times \ \ \ \ \ \ \ \ \ \ \ \ \ \ \ \ \ \ \ \ \ \ \ \\
\frac{P_d H_{d_{TX},d_{RX}}}{ \underset{s=1} {\overset{N}{\sum}} \rho_{d,s}P_s H_{s,d_{RX}} + \underset{s=1} {\overset{N}{\sum}} \underset{d^{'}\in \mathbb{M}\setminus \{d\}} {\overset{M}{\sum}} \rho_{d^{'},s} P_{{d^{'}}_{TX}} H_{{d^{'}}_{TX},d_{RX}} +N_d} )\}
\end{aligned}
\end{equation}
Subject to :
\begin{equation} \label{eq10}
\begin{aligned}
\frac{P_s (\alpha_s H_{s,B} +\beta_s H_{s,s}) }
{ \underset{d=1} {\overset{M}{\sum}} \rho_{d,s} P_d(\alpha_s H_{d_{TX},B} +\beta_s H_{d_{TX},s} )+N_s}\ \geq {\gamma_s}^{th},\  s=1..N
\end{aligned}
\end{equation}
\begin{equation} \label{eq11}
\begin{aligned}
\frac{P_d H_{d_{TX},d_{RX}}}{ \underset{s=1} {\overset{N}{\sum}} \rho_{d,s}P_s H_{s,d_{RX}} +\underset{s=1} {\overset{N}{\sum}} \underset{d^{'}\in \mathbb{M}\setminus \{d\}} {\overset{M}{\sum}} \rho_{d^{'},s} P_{{d^{'}}_{TX}} H_{{d^{'}}_{TX},d_{RX}} +N_d}\\ \geq {\gamma_d}^{th},\ \ \ \  d=1..M,\ \ \ \ \ \ \ \ \ \ \ \ \ \ \ \ \ \ \ \ \ \ \ \ \ \ \ \ \ \
\end{aligned}
\end{equation}
\begin{equation} \label{eq12}
\begin{aligned}
P_s \leq {P_s}^{max},\ s\in [1, N-V]\ and \\P_s \leq {P_d}^{max},\ s\in [N-V+1,N]
\end{aligned}
\end{equation}
\begin{equation} \label{eq13}
\begin{aligned}
P_d \leq {P_d}^{max}
\end{aligned}
\end{equation}
\begin{equation} \label{eq14}
\begin{aligned}
\underset{s=1} {\overset{N}{\sum}} \rho_{d,s} \in \{0,1\},\ d=1..M
\end{aligned}
\end{equation}
Here, the two first constraints are used to impose the SINR requirements for both UTs and DTs. The third and the fourth are used to ensure a limit for the maximal transmit power. The last constraint is adopted to guaranty that each DT can reuse one RB (that is equivalent to share the same RB of one UT).

Due to non-linear-integer formulation, the solution of this optimization problem is not straightforward and hard to find within short range time especially with large number of DTs. Therefore, we resort to a low complexity algorithm that manages the spectrum resources between the different users.

\section{Spectrum Resource Management Scheme}
The proposed algorithm for resource allocation is based on interference alleviation in the uplink. For that purpose, we define the total interference received at $d_{RX}$ by $
Interf_d= \underset{s=1} {\overset{N}{\sum}} \rho_{d,s}P_s H_{s,d_{RX}} +\underset{s=1} {\overset{N}{\sum}} \underset{d^{'}\in \mathbb{M}\setminus \{d\}} {\overset{M}{\sum}} \rho_{d^{'},s} P_{{d^{'}}_{TX}} H_{{d^{'}}_{TX},d_{RX}}$. To satisfy $\gamma_d$ defined in (\ref{eq11}), $Interf_d$ must not exceed the limit ${Interf_d}^{th}$ which is given as ${Interf_d}^{th}=\frac{P_d H_{d_{TX},d_{RX}}}{{\gamma_d}^{th}}-N_d$.\\
Likewise, we define $InterfB_s$ as the sum of interferences received at BS created by the set of DTs which reuse the RB of $UT_s$. Besides, we recall that $s=d^{*}$ if $(s>K)$. So, the interference received at ${d^{*}}_{RX}$ of $UT_s$ pair is included in $InterfB_s$. Thus, it is calculated as $\underset{d=1} {\overset{M}{\sum}} \rho_{d,s} P_d(\alpha_sH_{d_{TX},B}+\beta_sH_{d_{TX},s}) $. To satisfy $\gamma_s$  defined in (\ref{eq10}), $InterfB_s$ must not exceed ${InterfB_s}^{th}$ that corresponds to $ {InterfB_s}^{th}=(\alpha_s \frac{P_s H_{s,B}}{ {\gamma_s}^{th}}+\beta_s \frac{P_s H_{s,s}}{ {\gamma_s}^{th}})+N_s$. In what follows, $\mathbf{U}$ and $\mathbf{M^{'}}$ are intermediate sets used to refer to not yet allocated RBs and untreated DTs.

The scheduling mechanism, starts by treating each $DT_d$ from $\mathbf{M^{'}}$ apart. It considers all the possible links between $DT_d$ and UTs achieving $InterfB_s<{InterfB_s}^{th}$ then calculates $R_s+R_d$ with the assumption that $d^{'}=\emptyset , \ \forall d^{'}\neq d$. Afterwards, it chooses $UT_s$ that corresponds to $\mathrm{argmax}\{ R_d + R_s \}$ and adds $DT_d$ to $\Omega_s$. We refer to this step by cond1 in (\textbf{II}). Once the $\Omega$ are defined, the algorithm proceeds by checking the requirements in SINR for both DTs and UTs in order to manage the superposed interference for the users who reuse the same RBs (\textbf{III}). The resulting $\rho$ is the solution of the optimization problem (\textbf{IV}). A summary of the resource allocation and interference alleviation process is given in Alg.\ref{algo3}.
\begin{algorithm}
\DontPrintSemicolon
\textbf{I:} Initialize the network parameters: $\mathbf{U}=\mathbf{S},\mathbf{M}^{'}=\mathbf{M},$\\
Define the sets $\Omega=\{\Omega_1,\Omega_2,...,\Omega_N \}$ of possible assignment between UTs and DTs with $\Omega_i = \emptyset,\ i=1..N$\;
Define the sets $\rho= \{ \rho_1, \rho_2, ....,\rho_N\}$ of the optimal assignment with $\rho_i = \emptyset,\ i=1..N$\;
\textbf{II:} Fill $\Omega$ with the possible association between $\mathbf{M}^{'}$ and $\mathbf{U}$ according to cond1.\;
\Repeat{All elements in U are treated} {
\textbf{III:} Treat each $\mathbf{U}$ consecutively starting with the $UT_s$ who has the largest $\Omega_s$:\;
(1): Eliminate the DTs from $\Omega_s$ if their $Interf_d >{Interf_d}^{th}$.\;
(2): Each DT in $\Omega_s$ computes its own contribution in $InterfB_s$.\;
(3): Exclude from $\Omega_s$ the $DT_d$ with the greatest interference in the BS while $InterfB_s>{InterfB_s}^{th}$.\;
\textbf{IV:} Update:\;
(1): Mark $\rho_s=\Omega_s$ as the optimal assignment of $RB_s$.\;
(2): Eliminate $UT_s$ from $\mathbf{U}$.\;
(4): Apply \textbf{II} for all removed DTs.
}
\caption{Spectrum Resource Management paradigm\label{algo3}}
\end{algorithm}

\section{Numerical results}
In this section, simulation results are illustrated in order to assess the performance of our proposed algorithm. Details of the simulation parameters are given in Tab.\ref{SimulationParameters}

\begin{table}[!t]
\caption{Simulation Parameters} \label{SimulationParameters}
\centering
\begin{center}
\begin{tabular}{|M{5cm}|M{3cm}|}
\hline
\textbf{Parameter} &  \textbf{Value} \tabularnewline
\hline
    cell radius  & 500m \tabularnewline
\hline
   UT, DT min. close-in distance to BS & 35m  \tabularnewline
\hline
   Bandwidth per RB & 12 $\times$ 15 kHz $=$ 180 kHz \tabularnewline
\hline
   SINR requirement for each UTs ${\gamma_s}^{th}$  & 0dB \tabularnewline
\hline
   SINR requirement for each DTs ${\gamma_d}^{th}$  & 0dB \tabularnewline
\hline
   Noise power density & -174 dBm/Hz \tabularnewline
\hline
   UT total Tx. power $P_s$  & 30 dBm \tabularnewline
\hline
   DT total Tx. power $P_d$ & 10 dBm \tabularnewline
\hline
   Averaging window size $T_{p}$  & 500 \tabularnewline
\hline
    Path Loss LOS $(\mu,\nu,\sigma)$ parameters &  (61.4,2,5.8dB)\tabularnewline
\hline
   Path Loss N-LOS $(\mu,\nu,\sigma)$ parameters & (72,2.92,8.7dB)\tabularnewline
\hline
   Path loss probability for D2D links & $p1=0.8$ \tabularnewline
\hline
   Path loss probability for no D2D links & $p2=0.2$ \tabularnewline
\hline
Rician channel K factor & 5 \tabularnewline
\hline
Minimum required data per DT & 512Kb/s \tabularnewline
\hline
\hline
\end{tabular}
\end{center}
\end{table}
\begin{figure}
\centering
\includegraphics[width=8cm]{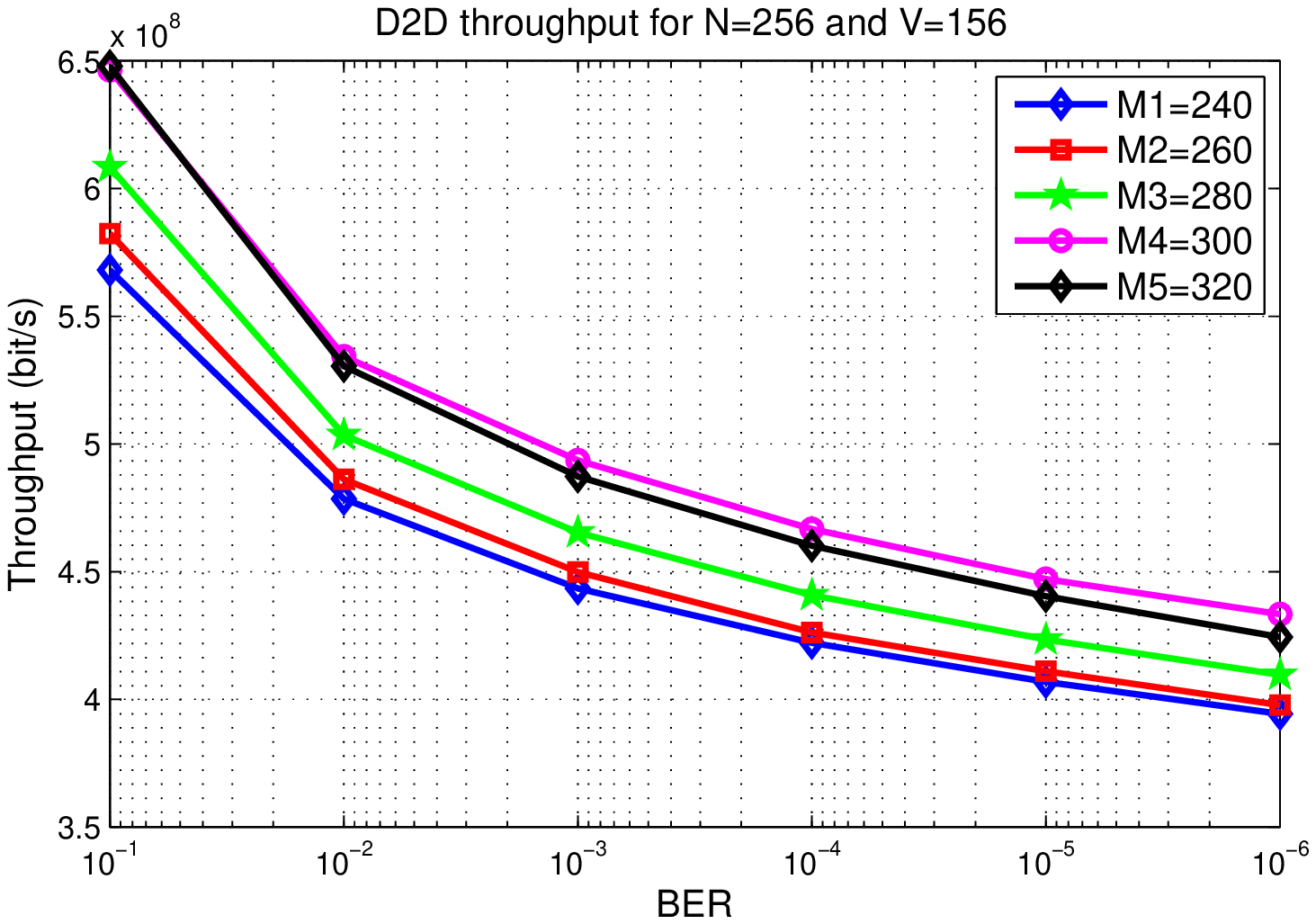}
\caption{D2D throughput with fixed number of UTs and variable DTs }\label{D2DthroughputvariableDTs}
\end{figure}
\begin{figure}
\centering
\includegraphics[width=8cm]{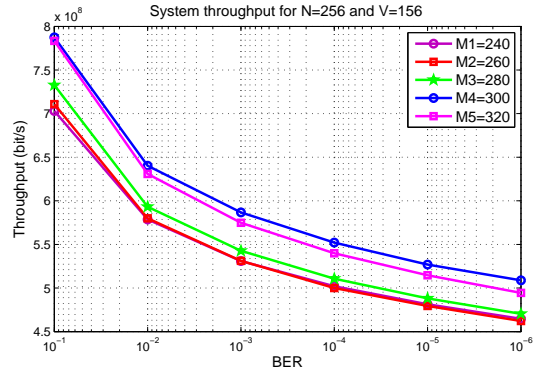}
\caption{System throughput with fixed number of UTs and variable DTs }\label{SystemThroughputVariableDTs}
\end{figure}
As an alternative interpretation of the results shown by Fig.\ref{D2DthroughputvariableDTs}, the rate achieved by the D2D communication increases steadily when the number of DTs goes up for a fixed number of UTs. This is reflected on the overall system performance, where the achievable rate of the  system has exactly the same shape as given in Fig.\ref{SystemThroughputVariableDTs}. For a maximum probability error equal to $10^{-1}$ the system rate rose by 0.08 Gbits/s from about 0.7 to 0.78  Gbits/s. This is primarily the result of the spatial reuse and proximity gain which improves consequently the overall system throughput. However, when the number of DTs becomes large, the case of $M=320$, the achievable rate for both graphs drops. This implies that eventually a very large number has a bad impact on the system performance that can be affected by the resulting interference. Moreover regardless of the number of users, the rate for the two graphs has been decreasing as long as the QoS defined by the BER is relaxed.\\
\begin{figure}
\centering
\includegraphics[width=8cm]{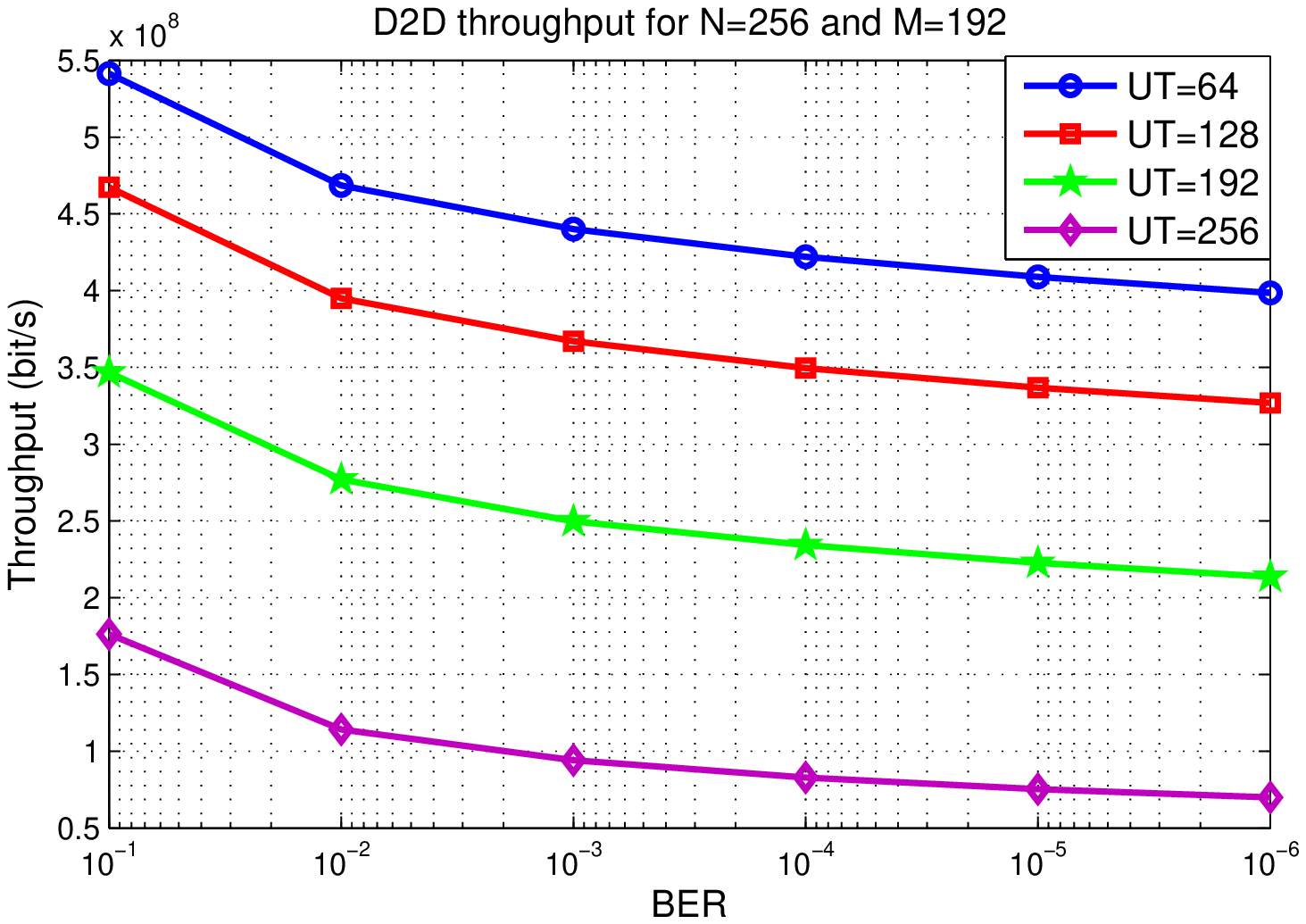}
\caption{D2D throughput with fixed number of DTs and variable UTs  }\label{D2DThroughputvariableUTs}
\end{figure}
\begin{figure}
\centering
\includegraphics[width=8cm]{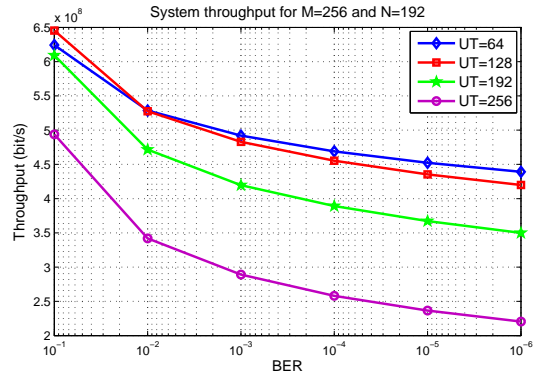}
\caption{System throughput with fixed number of DTs and variable UTs }\label{SystemThroughputVariableUTs}
\end{figure}
From Fig.\ref{D2DThroughputvariableUTs}, the D2D achievable rate decreases as long as the number of cellular users rises for a given number of DTs. This is a natural result, as the scheduling process of our proposed solution prioritizes the UTs since the RBs are firstly assigned to them and reused in the second order by DTs. Results from Fig.\ref{SystemThroughputVariableUTs} show that even the system throughput decreases with the increase of UTs number. This is due to the decrease in D2D data rate which affects the spatial reuse gain. It is also noticed the impact of the QoS defined by the BER, as the achievable rates drop with greater bit error probabilities.\\
\begin{figure}
\centering
\includegraphics[width=8cm]{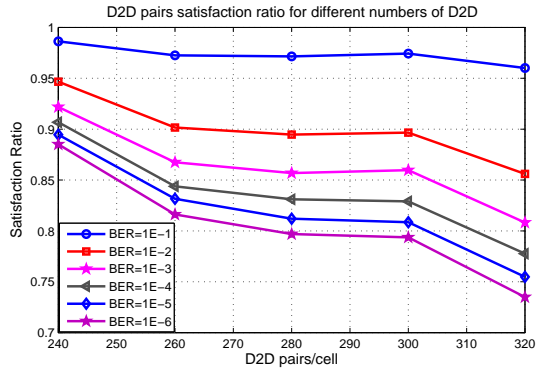}
\caption{Satisfaction ratio with fixed number of UTs and variable DTs }\label{D2DsatisfactionRatiovariableDTs}
\end{figure}
\begin{figure}
\centering
\includegraphics[width=8cm]{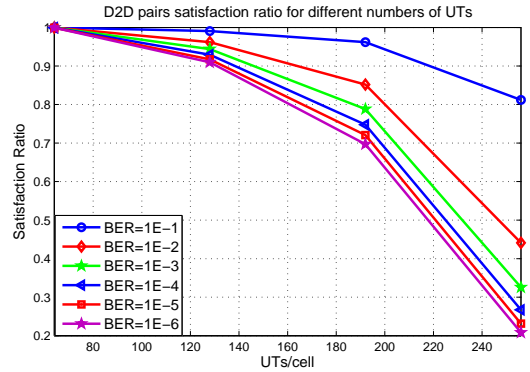}
\caption{Satisfaction ratio with fixed number of DTs and variable UTs }\label{D2DpairsSatisfactionvariableUTs}
\end{figure}
In what follows, we introduce the D2D satisfaction ratio (SR) as a new metric to assess the performance of the proposed scheduling scheme. The SR corresponds to the number of DTs whose data rate exceeds a minimum required level divided by their admitted number in the system. It is observed from Fig.\ref{D2DsatisfactionRatiovariableDTs} that the SR is about the same for $M=240$ till $M=300$. But when M exceeds 300, it decreases for all values of BER. This is due to the superposed interference in the system between the pairs that reuse the same RB. In Fig.\ref{D2DpairsSatisfactionvariableUTs}, $UT=64$ and $UT=128$ correspond to the totality or the half of DTs who are assigned individually RBs (i.e treated as UTs). Hence, their needs in spectrum resources are highly fulfilled compared to the other presented statistics. And besides that, when the QoS defined by the BER is high, the SR is improved.

These results are in great agreement with those treating the achievable data rate discussed above. They highlight how the spatial reuse and the multi-user reuse improve the network performance as long as the interference is well managed.
\section{Conclusion}
Throughout this paper, we considered the spectrum resource management in outdoor mmWave cell for the uplink of conventional and D2D communication. We aimed to optimize the system performance in terms of achievable throughput while achieving a compromise between the elevated number of admitted devices and the generated interference constraint. We provided a mathematical formulation of the optimization problem which falls in the mixed integer-real optimization scheme. To overcome its complexity, we proposed a heuristic algorithm and tested its efficiency through simulation results.


\bibliographystyle{IEEEtran}
\bibliography{article3bib}
\end{document}